\documentclass[12pt]{iopart}
\usepackage{iopams}
\usepackage{graphicx,
longtable}
\begin{document}
\title{Overconstrained estimates of neutrinoless double beta decay within the QRPA}
%
\author{ A.~Faessler$^1$, G.L.~Fogli$^{2,3}$, E.~Lisi$^{3}$, V.~Rodin$^1$, A.M.~Rotunno$^{2,3}$, F.~\v{S}imkovic$^1$}
\address{$^1$ Institute of Theoretical Physics,
				University of Tuebingen,\\
                72076 Tuebingen, Germany}
\address{ $^2$   Dipartimento Interateneo di Fisica ``Michelangelo Merlin,''\\
                Via Orabona 4, 70126 Bari, Italy}
\address{$^3$   Istituto Nazionale di Fisica Nucleare, Sezione di Bari,\\
                Via Orabona 4, 70126 Bari, Italy}
\ead{vadim.rodin@uni-tuebingen.de}
\begin{abstract}
Estimates of nuclear matrix elements for neutrinoless double beta decay ($0\nu2\beta$) 
based on the quasiparticle random phase approximations (QRPA) are affected by 
theoretical uncertainties, which can be substantially reduced by fixing the 
unknown strength parameter $g_{pp}$ of the residual particle-particle interaction 
through one experimental constraint --- most notably through the two-neutrino double 
beta decay ($2\nu2\beta$) lifetime. However, it has been noted that the $g_{pp}$ 
adjustment via $2\nu2\beta$ data may bring QRPA models in disagreement
with independent data on electron capture (EC) and single beta decay ($\beta^-$) 
lifetimes. Actually, in two nuclei of interest for $0\nu2\beta$ decay
($^{100}$Mo and $^{116}$Cd), for which all such data are available, 
we show that the disagreement vanishes, provided that the axial vector coupling 
$g_A$ is treated as a free parameter, with allowance for $g_A<1$ (``strong quenching''). 
Three independent lifetime data ($2\nu2\beta$, EC, $\beta^-$) are then accurately 
reproduced by means of two free parameters $(g_{pp},\,g_A)$, resulting in an 
overconstrained parameter space. In addition, the sign of the $2\nu2\beta$ matrix element
$M^{2\nu}$ is unambiguously selected $(M^{2\nu}>0)$ by the combination of all data. 
We discuss quantitatively, in each of the two nuclei, these phenomenological 
constraints and their consequences  for QRPA estimates of the $0\nu2\beta$ matrix 
elements and of their uncertainties. 
\end{abstract}
\medskip
\submitto{\JPG}
\maketitle

\section{Introduction \label{SecI}}

The new paradigm of massive and mixed neutrinos, emerging from the recent evidence 
for neutrino flavor oscillations \cite{Revi,Rev2,Rev3,Rev4}, is still incomplete in
several aspects. In particular, the nature of the neutrino fields (Dirac or Majorana)
\cite{PeBi}  remains undetermined, the amount of CP violation in the neutrino sector (if any) 
is unconstrained, and the absolute neutrino masses---as well as their ordering---are not 
yet known. The process of neutrinoless double beta decay ($0\nu2\beta$),
\begin{equation}
(Z, A)\to (Z+2, A) + 2e^- \ \ \ (0\nu2\beta)\ ,
\end{equation}
bears on all these issues and, thus, is a major research topic 
in current experimental and theoretical neutrino physics \cite{Fa98,Su98,Vo06,Av07}.
The claimed observation of $0\nu2\beta$ decay in $^{76}$Ge with lifetime 
$T^{0\nu}_{1/2}\simeq 2.2\times 10^{25}$~y \cite{Klap}, and the projects aimed at 
its independent (dis)confirmation \cite{Av07}, have also given new impetus to the field.

In general, barring contributions different from light Majorana neutrino exchange, 
the inverse $0\nu2\beta$ lifetime in a given nucleus is the product of three factors,
\begin{equation}
\label{3fact}
\left(T^{0\nu}_{1/2}\right)^{-1}=G^{0\nu}\,\left|M^{0\nu}\right|^2\,m_{\beta\beta}^2\ ,
\end{equation}
where $G^{0\nu}$ is a calculable phase space factor, $M^{0\nu}$ is the $0\nu2\beta$ nuclear
matrix element, and $m_{\beta\beta}$ is the (nucleus-independent)
``effective Majorana neutrino mass'' which, in standard notation \cite{PDGr}, reads
\begin{equation}
m_{\beta\beta}=\left|\sum_{1=1}^3 m_i\,U^2_{ei}\right|\ ,
\end{equation}
$m_i$ and $U_{ei}$ being the neutrino masses and the $\nu_e$ mixing matrix elements,
respectively.

The calculation of the matrix element $M^{0\nu}$~in Eq.~({\ref{3fact}}) for a candidate $0\nu2\beta$
nucleus is notoriously difficult. It requires the detailed description of a second-order 
weak decay from a double-even ``mother'' nucleus $(Z,\,A)$ to a double-even
``daughter''  nucleus $(Z+2,\,A)$ via virtual states (with any multipolarity $J^\pi$)
of the so-called ``intermediate'' nucleus $(Z+1,\,A)$. The decay can proceed through both Fermi (F)
and Gamow-Teller (GT) transitions, plus a small tensor (T) contribution,
\begin{equation}
M^{0\nu} = M^{0\nu}_\mathrm{GT}+M^{0\nu}_\mathrm{T}-\displaystyle\frac{M_\mathrm{F}^{0\nu}}{g^2_A}\ ,
\end{equation}
and detailed nuclear models are required to estimate the separate components
$M^{0\nu}_\mathrm{X}$ ($\mathrm{X}=\mathrm{F}$, GT, T). In the above expression,
 $g_A$ is the effective axial coupling in nuclear matter, not necessarily equal 
to its ``bare'' free-nucleon value $g_A\simeq1.25$ \cite{Beta}.

Modern calculations of $0\nu2\beta$ matrix elements are usually  
performed within either the quasiparticle random phase approximation (QRPA) \cite{Ro07,Ko07} 
or the nuclear shell model (NSM)  \cite{Ca05} and their variants, sometimes with large differences
among the results. We remind that the QRPA basis of nuclear many-particle configurations,
on which the residual particle-hole and particle-particle interaction is diagonalized to build the 
nuclear excitations, is limited to iterations of two-quasiparticle ones (reducing to the particle-hole 
configurations when the pairing interaction is switched off); for details, see, e.g.~\cite{Fa98,Su98}.
The advantage of the QRPA as compared to the NSM is that one can include essentially unlimited sets 
of single-particle states, even those forming the continuum of the positive-energy ones within the 
continuum-QRPA~\cite{Ro07a}.

Painstaking but steady progress in both the QRPA and the NSM approaches is gradually leading to a 
better understanding---and to a reduction---of the differences among their results \cite{Av07}. However, even 
in the most refined approaches, the estimates of $M^{0\nu}$ remain affected by various  
uncertainties, whose reduction is of paramount importance for both theory and experiment.  
Indeed, uncertainties in $M^{0\nu}$ propagate to the extracted value
of (or limit on) $m_{\beta\beta}$ via Eq.~(\ref{3fact}), and affect directly the 
design of $0\nu2\beta$ experiments (in particular the detector size and the choice of the nucleus) 
needed to reach a given target sensitivity to $m_{\beta\beta}$ \cite{Av07}. Among the sources
of uncertainties one can quote: 
(1) inherent approximations and simplifications of the theory; 
(2) existence of free or adjustable model parameters; 
(3) problematic description of the strong short-range repulsive interaction between nucleons; and 
(4) uncertainties in the value of $g_A$.

The latter problem arises from the significant reduction (``quenching'') of the strength  
observed in nuclear GT transitions (see, e.g.,  
\cite{Ost91}), which still lacks a clear experimental quantification and theoretical understanding. 
Two possible physical origins of the quenching  have been discussed, one due to the $\Delta$-isobar admixture 
in the nuclear wave function~\cite{Bor81} and another one 
due to the shift of the Gamow-Teller strength to higher excitation energies induced
by short range tensor correlations \cite{Ber82}.
In the absence of a better prescription, the effect of quenching (in either 
QRPA or NSM calculations) is often simply evaluated by replacing the bare value $g_A\simeq 1.25$ 
with an empirical, quenched value $g_A\simeq 1$ \cite{El04}. However, there is no {\em a priori\/} 
reason to exclude values $g_A\lesssim 1$, which have indeed sometimes been advocated, especially 
within the NSM approach \cite{Ca05,Ejir}.

In this context, we present a novel approach towards data-driven constraints on $M^{0\nu}$ calculations,
assuming the possibility of strong quenching ($g_A<1$) within the QRPA. 
This unconventional hypothesis makes theory and data agree in a number of cases, where previous
attempts have systematically failed. Therefore, we think that our approach may lead to a fruitful 
discussion and a fresh look at the whole problem of quenching, from both the theoretical and the experimental
viewpoint.  We stress, however, that we simply treat $g_A<1$ as a phenomenological possibility in this work,
without any attempt to elaborate theoretical interpretations of the $g_A$ values
emerging from the data analysis.

Our work is structured as follows. In Sec.~\ref{SecII} we discuss the experimental data which can be used to benchmark 
the QRPA model. We adopt a selected data set, including the measured lifetimes
of two-neutrino double beta decay, electron capture, and single beta decay for two nuclei, $^{100}$Mo and $^{116}$Cd,
which are of interest for searching $0\nu2\beta$ decay. In Sec.~\ref{SecIII} we compare these data with the corresponding
QRPA results, assuming standard quenching ($g_A=1$) or no quenching ($g_A=1.25$). We face then 
the well-known problem that
the theory cannot match two or more data at the same time, for any given value of the
so-called particle-particle strength parameter $g_{pp}$. In Sec.~\ref{SecIV} we show that this
problem can be phenomenologically removed if strong quenching ($g_A<1$) is allowed. In this
case, the two parameters $(g_{pp},\,g_A)$ are overconstrained by three independent data in each of
the two chosen nuclei, as shown in Sec.~\ref{SecV}. In Sec.~\ref{SecVI} we propagate the estimated 
$(g_{pp},\,g_A)$ uncertainties to the calculation of $0\nu2\beta$ matrix elements and lifetimes, with and 
without the effects
of short-range repulsive interactions. Finally, we summarize our results and discuss future perspectives 
in Sec.~\ref{SecVII}.

\section{Experimental benchmarks \label{SecII}}

In order to reduce the theoretical uncertainties, any nuclear model used in $0\nu2\beta$ 
calculations should be benchmarked by as many weak-interaction data \cite{ATOM} as possible.
Relevant weak processes are listed in Eqs.~(\ref{2beta})--(\ref{muC}) below.

Two-neutrino double beta decay ($2\nu2\beta$), 
\begin{equation}
\label{2beta}
(Z,A)\to (Z+2,A)+ 2e^-+2\overline\nu_e \ \ \ (2\nu2\beta)\ ,
\end{equation}
is a second-order weak process ($|\Delta Z|=2$) which probes the same mother and daughter
nuclei as $0\nu2\beta$ decay. It has been observed in several nuclei, 
thus providing a particularly important benchmark. 
Indeed, it was extensively demonstrated in~\cite{Ro07} that the spread of QRPA 
calculations  can be significantly reduced by constraining the nuclear model
with the corresponding experimental $2\nu2\beta$ decay lifetime (see \cite{St92} for earlier
attempts). The $2\nu2\beta$ data help to fix an important free model parameter, namely,
the strength $g_{pp}$ of the residual particle-particle interaction~\cite{Ch83,Vo86}, 
and thus to ``calibrate'' the QRPA estimates of $M^{0\nu}$. Despite the fact
that the $2\nu2\beta$ decay process probes only a subset of the  intermediate
states relevant for $0\nu2\beta$ decay (i.e., only those with $J^\pi=1^+$, via GT transitions), it is just
the $1^+$ contribution to the total $0\nu2\beta$ matrix element that reveals a pronounced 
sensitivity to $g_{pp}$, in contrast to the other multipole contributions~\cite{Sim07}. 
This observation justifies the aforementioned fitting procedure employed in~\cite{Ro07}.

First-order  weak processes ($|\Delta Z|=1$) related to $0\nu2\beta$ decay 
can probe, in usual jargon, either the ``first leg'' of the decay
(from the mother nucleus to the intermediate one) or its ``second leg'' (from the
intermediate nucleus to the daughter one). Relevant examples for the first leg  
include the electron capture (EC) from a bound state ($e^-_b$),
\begin{equation}
\label{EC}
e^-_b+(Z+1,A)\to (Z,A)+\nu_e\ \ \ (\mathrm{EC})\ ,
\end{equation}
and the charge-exchange reaction via $(^3\mathrm{He},t)$,
\begin{equation}
^3\mathrm{He} + (Z,A)\to (Z+1,A)+{}^3\mathrm{H}\ \ \ (^3\mathrm{He},t)\ ,
\end{equation}
as well as via $(p,n)$,
\begin{equation}
p+(Z,A)\to (Z+1,A)+n\ \ \ (p,n)\ .
\end{equation}
The second leg is instead probed by the $\beta^-$ decay,
\begin{equation}
\label{beta}
(Z+1,A)\to(Z+2,A)+e^-+\overline\nu_e \ \ \ (\beta^-)\ ,
\end{equation}
by the charge-exchange $(d,{}^2\mathrm{He})$ reaction,
\begin{equation}
^2\mathrm{H} + (Z+2,A)\to (Z+1,A)+{}^2\mathrm{He}\ \ \ (d,{}^2\mathrm{He})\ ,
\end{equation}
and by ordinary muon capture ($\mu$C),
\begin{equation}
\label{muC}
\mu^-+(Z+2,A)\to(Z+1,A)+\nu_\mu \ \ \ (\mu\mathrm{C})\ .
\end{equation}
See also \cite{Zu05} for a recent discussion of these and other possible weak processes, 
including future (anti)neutrino-nucleus charged-current reactions
at low energy \cite{Vo05}. 
Clearly, any of the above first-order weak processes could be used to
set useful constraints on the nuclear model. Indeed, using $\beta^-$ decay has been advocated as  
an alternative to $2\nu2\beta$ decay for fixing the $g_{pp}$ parameter in QRPA 
\cite{Ci05,Su05,SuBe}; $\mu$C data might be similarly used in the near future \cite{Ko06}. 
However, one should be aware that these data are currently more sparse than for $2\nu2\beta$ 
decay and, sometimes, have inherent problems or limitations, as discussed below.

\Table{\label{tab:overview}  Compilation of experimental references for nine nuclear systems ($A$) of interest 
in $0\nu2\beta$ decay  (``$Z$'', ``$Z+1$'', ``$Z+2$'' denote ``mother'', ``intermediate'', ``daughter'' nuclei, respectively). 
The entries refer to $2\nu2\beta$ decay data ($|\Delta Z|=2$) as well as to processes probing
the so-called first and second leg ($|\Delta Z|=1$). For $\beta^-$ data, only decays from 
$J^\pi=1^+$ states are considered. For muon capture ($\mu$C), the data in \cite{Suzu} actually refer
to natural isotopic mixture of the $Z+2$ nucleus. See also: \cite{TiEC,Frek} for proposed EC measurements
at $A=76$, 82, 100, 116, and 128; \cite{Frek} for proposed ($^3$He,$\,t$) measurements at $A=76$ and
preliminary ($d$,$\,{}^2$He) data at $A=76$ and 96; \cite{MuCA} for preliminary $\mu$C data at $A=76$, 82, and 150.
}
\br
\multicolumn{4}{c}{Nuclei} & 
$|\Delta Z|=2~~~$ & 
\multicolumn{3}{c}{$|\Delta Z|=1$, first leg} & 
\multicolumn{3}{c}{$|\Delta Z|=1$, second leg} \\[1mm]
\mr
$A$ &$Z$&$Z+1$&$Z+2$&$2\nu2\beta$& EC               & ($^3$He,$\,t$)& $(p,\,n)$ & $\beta^-$   & ($d$,$\,{}^2$He)&$\mu$C  \\[1mm]
\mr
76  & Ge & As & Se & \cite{Bara} & 			       &             & \cite{Made} &             &             & \cite{Suzu} \\
82  & Se & Br & Kr & \cite{Bara} &                  &             & \cite{Made} &             &             &             \\
96  & Zr & Nb & Mo & \cite{Bara} &                  &             &             &             &             & \cite{Suzu} \\
100 & Mo & Tc & Ru & \cite{Bara} & \cite{Garc,Wilk} & \cite{Akim} &             & \cite{ENSD} &             &             \\
116 & Cd & In & Sn & \cite{Bara} & \cite{Bhat}      & \cite{Akim} & \cite{Sasa} & \cite{ENSD} & \cite{Rake} & \cite{Suzu} \\
128 & Te & I  & Xe & \cite{Bara} & \cite{ENSD}      &             & \cite{Made} & \cite{ENSD} &             &             \\
130 & Te & I  & Xe & \cite{Bara} &                  &             & \cite{Made} &             &             &             \\
136 & Xe & Cs & Ba & \cite{Bara} &                  &             &             &             &             & \cite{Suzu} \\
150 & Nd & Pm & Sm & \cite{Bara} &                  &             &             &             &             & \cite{Suzu} \\[1mm]
\br
\endTable
%

Table~\ref{tab:overview} shows the current experimental status of the seven processes
listed in Eqs.~(\ref{2beta}--\ref{muC}), for nine nuclei  of interest for $0\nu2\beta$ decay searches.
Data on $2\nu2\beta$ decay lifetimes exist for all these nuclei \cite{Bara}. 
Lifetimes for EC and $\beta^-$ decay have been measured only in three cases, $A=100$, 116 and 128
(with $J^\pi=1^+$ states for the intermediate nucleus) \cite{Garc,Wilk,Bhat,ENSD}. In one case ($A=100$), the most
recent EC datum \cite{Wilk} appears to be in conflict with the older one \cite{Garc}. Data on the charge-exchange scattering
processes are also sparse. Available $\mu$C data \cite{Suzu} are not particularly constraining, since they refer 
to the natural isotopic mixture containing the daughter nucleus; see however \cite{Zi06} for a comparison of QRPA 
calculations with $\mu$C data, and \cite{MuCA} for preliminary $\mu$C data in unmixed $A=76,$ 82, and 150 daughter
nuclei. Charge-exchange reactions involve analyses of spectral data which are, in general, 
more difficult to be interpreted and modeled than decay lifetimes \cite{Ze06,Am08}.
Data for $(^3\mathrm{He},t)$ exchange are available only for $A=100$ and 116 \cite{Akim}. In the latter case, 
the measured GT strength is in conflict with the one derived from EC \cite{Bhat}. Data for  $(d,{}^2\mathrm{He})$ 
exchange and $A=116$ are reported in \cite{Rake}, where the GT strength distribution
is, however, normalized to the reference $\beta^-$ one \cite{ENSD} at small excitation energy,
and thus it does not provide an entirely independent constraint.  The $(p,n)$ reaction has been instead 
studied in several nuclei \cite{Made,Sasa}, with emphasis on the GT strength
distribution (rather than on its normalization). For $A=116$, it should be noted that the recent 
$(p,n)$ data in \cite{Sasa} disagree with the $(^3\mathrm{He},t)$ data in \cite{Akim}, and are only
in rough agreement with the EC data in \cite{Bhat}. 

Clearly, new and dedicated measurements are needed, both to solve the mentioned experimental discrepancies
and to fill the missing entries in Tab.~\ref{tab:overview} \cite{Zu05,TiEC,Frek}. In the meantime,
one needs to select a (hopefully consistent) data set, in order to perform a meaningful comparison with theoretical
calculations.

In this work we adopt the following approach: we ignore current data from the charge-exchange scattering processes
(which, in several cases, either disagree with each other, or have no independent normalization, or 
provide poor constraints for our purposes), and we choose only those data which involve 
half-life measurements (rather than complex spectral analyses), namely, 
$2\nu2\beta$, EC, and $\beta^-$ decay. Our investigation is then restricted to two nuclear systems for which 
all such data exist, namely, $A=100$ and 116, which we shall often denote by the name of the 
``mother'' nucleus ($^{100}$Mo and $^{116}$Cd, respectively). 
For $A=100$, we discard the old EC datum, $\log ft(\mathrm{EC}) \simeq 4.45$ \cite{Garc}, in favor of the new 
(albeit unpublished) one \cite{Wilk}. 
Table~\ref{tab:inputdata} shows the corresponding input data that will be used in our analysis, 
in terms of $\log T/\mathrm{y}$ (for $0\nu2\beta$) and of $\log ft$ (for EC and $\beta^-$), 
where $f$ is the usual nuclear Fermi function.  (Throughout this paper, $\log\equiv\log_{10}$.) 

Although the ($2\nu2\beta$, EC, $\beta^-$) data are available also for A=128 (see Table~1), this nuclear system
is left out of the consideration in the present work since the final nucleus $^{128}$Xe is rather strongly 
deformed. The change in the deformation from an almost spherical $^{128}$Te to a rather well
deformed $^{128}$Xe ($\beta=0.18$~\cite{Ram0}) cannot be reliably treated within the spherical QRPA employed here. 
Importance of such an effect has been demonstrated in Refs.~\cite{Si04,Alv04} for the case of the $2\nu2\beta$ decay
using the deformed QRPA with schematic forces. 

\ \hskip-5cm \Table{\label{tab:inputdata} Experimental input. Half-life data (with $1\sigma$ experimental errors) 
for $2\nu2\beta$, EC, and $\beta^{-}$ decay in $^{100}$Mo and $^{116}$Cd. All logarithms are in base 10.}
\br
Nucleus     &$\log(T^{2\nu}_{1/2}/\mathrm{y})\pm\sigma_\mathrm{exp}$
                                            & Ref.      &$\log ft(\mathrm{EC})\pm\sigma_\mathrm{exp}$
                                                                               & Ref.      & $\log ft(\beta^-)\pm\sigma_\mathrm{expt}$
                                                                                                                        & Ref. \\[1mm]
\mr
$^{100}$Mo&$18.85\pm0.03$                   &\cite{Bara}&$3.96^{+0.11}_{-0.09}$&\cite{Wilk}%
																						 & $4.60\pm0.01$     &\cite{ENSD} \\[1mm]
$^{116}$Cd&$19.48\pm0.03$                   &\cite{Bara}&$4.39^{+0.10}_{-0.15}$&\cite{Bhat}& $4.662\pm0.005$   &\cite{ENSD}
\\
\br
\endTable

\section{Data versus theory with standard or no quenching $\mathbf{(1\leq g_A \leq 1.25)}$ \label{SecIII}}

In the context of the QRPA, it has been convincingly shown in \cite{Ro07}
that the spread of theoretical calculations can be significantly reduced, in each
of the nine nuclei in Tab.~\ref{tab:overview}, by fixing $g_{pp}$ in such a way as to reproduce the 
measured $2\nu2\beta$ lifetimes.%
\footnote{It is worth noticing that, in general, the effective values of 
both $g_{pp}$ and $g_A$ may change in different nuclei.}
 This approach has however been questioned in \cite{Ci05,Su05}, since the 
fitted value of $g_{pp}$ appears to underestimate (overestimate) the EC ($\beta^-$) lifetime by a large factor, 
as compared with experimental data. The alternative choice of fitting $g_{pp}$ by reproducing, e.g., the
$\beta^-$ decay lifetime \cite{Su05,SuBe}, merely shifts the problem to other data (e.g., to the
$2\nu2\beta$ or EC lifetimes) which are no longer correctly reproduced; see also \cite{Gr92} for
early examples of such a conflict. It is worth noticing that, in the related
literature \cite{Ro07,SuBe,Ko07}, $g_A$ has been taken in the range $1\lesssim g_A \lesssim 1.25$, i.e., 
between standard quenching ($g_A\simeq 1$) and no quenching ($g_A\simeq 1.25$). Within such range, the 
problem of fitting two or more data (among  $2\nu2\beta,\,\mathrm{EC},\,\beta^-$)
appears to be basically unsolved in the QRPA. Before discussing in the detail this problem 
in Secs.~\ref{SecIIIA}--\ref{SecIIIB} below, we recall a few essential features of the QRPA.

The $(2\nu2\beta,\,\mathrm{EC},\,\beta^-)$ processes occur through GT transitions, either
at first order in $g_A$ (for EC, $\beta^-$) or at second order in $g_A$ (for $2\nu2\beta$).
Therefore, theoretical estimates of the associated (logarithmic) lifetimes  need to be performed 
only for $g_A=1$, and can then be scaled for $g_A\neq 1$ as:
\begin{eqnarray}
\log (ft) &\to& \log (ft/g_A^2)\ \  \mathrm{for\ EC\ and\ }\beta^-\ ,\label{ga1}\\
\log (T_{1/2}^{2\nu}) &\to& \log (T_{1/2}^{2\nu}/g_A^4) \ \  \mathrm{for\ }2\nu2\beta\ .\label{ga2}
\end{eqnarray}

Within this work, QRPA calculations of the above lifetimes have been performed both in large basis (l.b., default
choice) and in small basis (s.b.).  The small basis consists of 13 single-particle levels (oscillator shells 
$N=3$ and 4, plus the $f + h$ orbits from $N=5$), while the large basis contains 21 levels 
(all states from shells $N=1,\dots,5$),
in accordance with the choice of~\cite{Ro03,Ro07}. The small set corresponds
to $1 \hbar\omega$ particle-hole excitations, and the large one to about $4 \hbar\omega$ excitations.

An important output of QRPA calculations is the $2\nu2\beta$ matrix element $M^{2\nu}$, whose modulus is
probed by the observable $T_{1/2}^{2\nu}$ according to 
\begin{equation}
\frac{1}{T_{1/2}^{2\nu}} = G^{2\nu} \left(\frac{g_A}{1.25}\right)^4 \left|M^{2\nu}\right|^2 \ ,
\end{equation}
where $G^{2\nu}$ is a calculable phase space factor, and the bare value of $g_A$ (1.25) is explicitly 
factorized out to make contact with previous notation \cite{Ro07}. 
In QRPA calculations, $M^{2\nu}$ typically starts positive for $g_{pp}\ll 1$, then decreases 
and eventually changes sign as $g_{pp}$ increases. The critical value $g^*_{pp}$ where $M^{2\nu}=0$ marks 
an infinite lifetime, $\log T^{2\nu}_{1/2}\to \infty$. It turns out that $\log ft(\mathrm{EC})$ is continuous 
across $g^*_{pp}$, while $\log ft(\beta^-)$ diverges locally. For $g_{pp}$ increasing slightly beyond this
critical point, the calculated energy of the first excited state $E_1$  decreases and eventually
vanishes, inducing a breakdown (the so-called ``collapse'') of the QRPA solution.
QRPA calculations become thus less reliable in the vicinity of the critical and collapse points.

\begin{figure}[t]
\begin{center}
\includegraphics[scale=1.1]{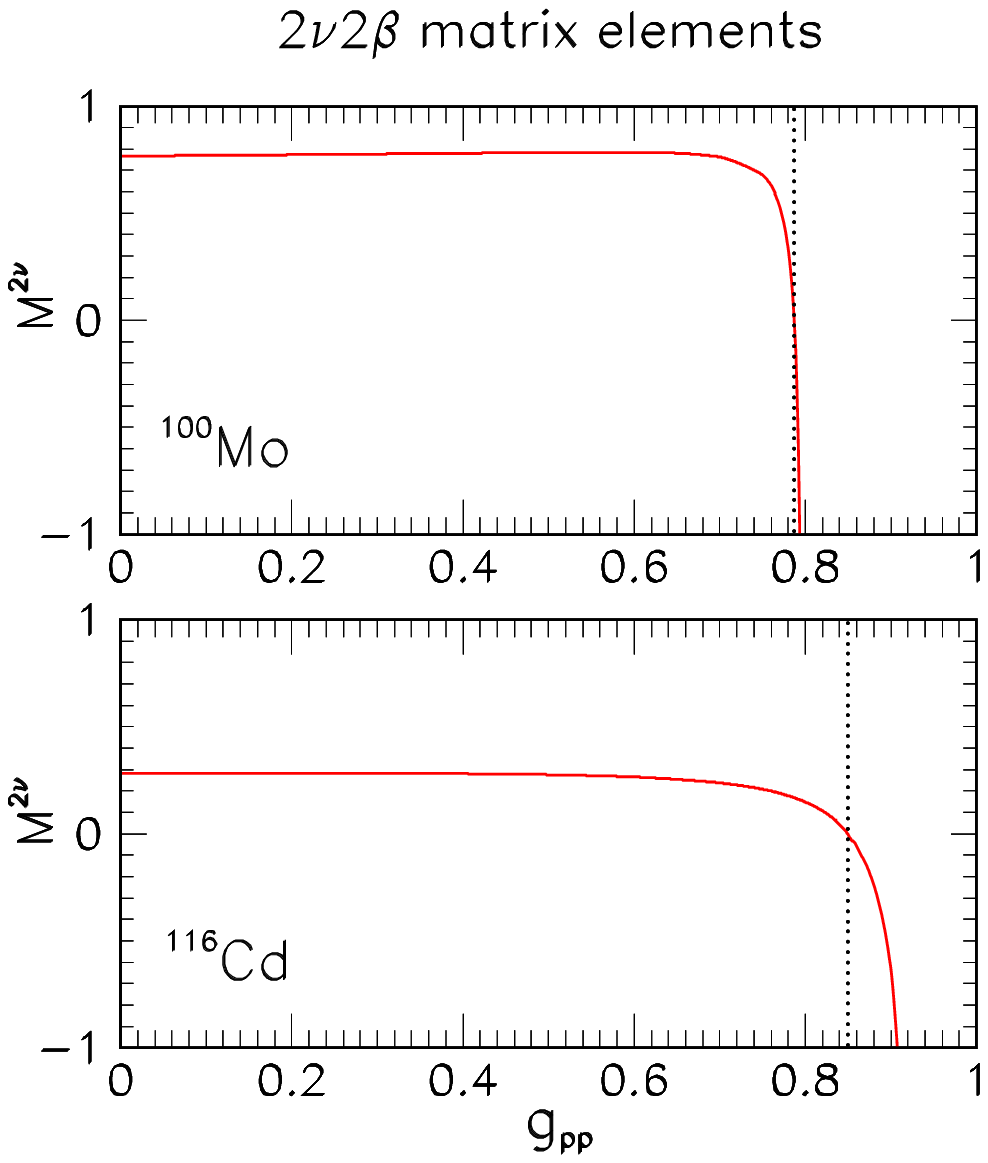}
\end{center}
\caption{ \label{f01} Matrix elements for $2\nu2\beta$ decay in the QRPA (solid curves) 
as a function of $g_{pp}$, for $^{100}$Mo and $^{116}$Cd. In each panel,
the vertical dotted line marks the critical $g_{pp}$ value where $M_{2\nu}=0$. Calculations
refer to the large basis.}
\end{figure}

Figure~\ref{f01} shows the matrix element $M^{2\nu}$ as a function of $g_{pp}$ for each of the 
two reference nuclei, in large basis. Similar results are found for small basis 
(not shown). In each panel, a vertical dotted line marks the critical value
$g^*_{pp}$ where $M^{2\nu}$ flips its sign. The value of $M^{2\nu}$ drops rapidly for $g_{pp}>g_{pp}^*$, 
and the QRPA collapse is eventually reached.  Both positive and negative values 
of $M^{2\nu}$ may be phenomenologically acceptable in principle, although
theoretical arguments suggest that $M^{2\nu}>0$ \cite{Ro07}. Determining the sign of $M^{2\nu}$ is thus a 
relevant check of the theory.

The QRPA estimates of $M^{2\nu}$, as well as those of the $2\nu2\beta,\,\mathrm{EC}$, and $\beta^-$ lifetimes,
are affected by various sources of uncertainties. In Sec.~\ref{SecV} we shall deal
with the uncertainties related to the ({\em a priori\/} unknown) values of $g_{pp}$ and $g_A$, and to the size of the basis.
However, even if $g_{pp}$ and $g_A$ were perfectly known and the basis size were irrelevant, the 
approximation inherent to the QRPA approach would introduce further theoretical errors on each 
estimated lifetime. The assessment of these errors is obviously difficult and, to some extent, even
arbitrary---but it is necessary to gauge the (dis)agreement between theoretical estimates and data.
Our educated guess for the {\em extra\/} theoretical uncertainties   
(besides those related to $g_{pp}$, to $g_A$, and to the basis size) is $\sim 20\%$ 
for both the EC and $2\nu2\beta$ lifetimes, and $\sim 40\%$ for the $\beta^-$ lifetime. 
In the latter case, a larger {\em relative\/} error is assumed, due to the smaller (by a factor 2--3) 
calculated values of the corresponding
matrix element as compared with the ones for the EC.
Accordingly, we attach the following ($\pm 1\sigma$) 
theoretical errors $\sigma_\mathrm{th}$ 
to each logarithmic lifetime, for any fixed values of $(g_{pp},\,g_A)$ in any basis: 
\begin{eqnarray}
\label{unc1}\log (T^{2\nu}_{1/2}/\mathrm{y}): 	&& \sigma_\mathrm{th}=\pm 0.08\ ,\label{error1}\\
\label{unc2}\log ft(\mathrm{EC}): 			&& \sigma_\mathrm{th}=\pm 0.08\ ,\label{error2}\\
\label{unc3}\log ft(\mathrm{\beta^-}): 		&& \sigma_\mathrm{th}=\pm 0.15\ .\label{error3}
\end{eqnarray}

In the next two subsections we shall compare the data in Tab.~\ref{tab:inputdata} 
with the corresponding QRPA estimates for $g_A=1$.
It will be shown that, in none of the two reference nuclei, the QRPA results can be really
made consistent with more than one datum at a time, within the quoted experimental and theoretical
uncertainties. Moreover, it will become evident that higher $g_A$ values (e.g., $g_A=1.25$) can
only worsen the situation.

\subsection{$^{100}$Mo data versus QRPA ($ {g_A=1}$) \label{SecIIIA}}

\begin{figure}[t]
\begin{center}
\includegraphics[scale=1.1]{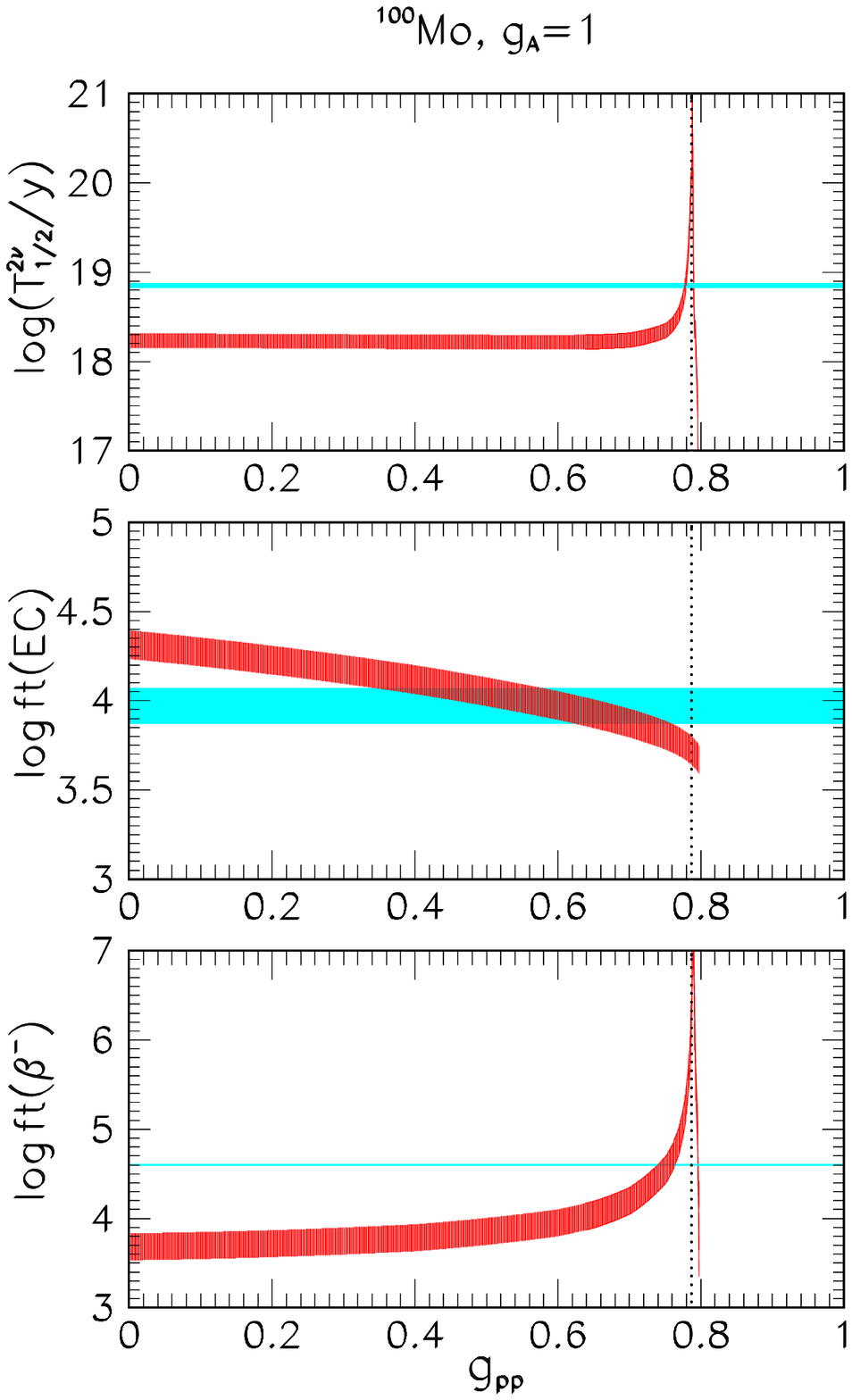}
\end{center}
\caption{ \label{f02} Lifetimes for the $2\nu2\beta$, EC, and $\beta^-$ decay in
$^{100}$Mo. Horizontal bands: experimental data. Curved bands: the QRPA results as 
a function of $g_{pp}$, for fixed $g_A=1$, in large basis. The vertical width of 
the bands corresponds to $\pm1\sigma$ uncertainties.}
\end{figure}

Figure~\ref{f02} illustrates the comparison between $^{100}$Mo data and theoretical predictions
for standard quenching ($g_A=1$) in large basis, as a function of $g_{pp}$. The upper, middle, and lower
panels refer to the $2\nu2\beta$, EC, and $\beta^-$ logarithmic lifetimes, respectively.
In each panel, the horizontal band represents the experimental
datum at $\pm 1\sigma$ (as taken from Tab.~\ref{tab:inputdata}), while the curved band represents the QRPA
results, with $\pm 1\sigma$ theoretical spread as in Eqs.~(\ref{error1}--\ref{error3}).
Vertical dotted lines mark the critical value $g_{pp}^*$ which separate the left, positive branch 
$(M^{2\nu}>0)$ from the right, negative branch $(M^{2\nu}<0)$. 
The preferred $g_{pp}$ ranges---where the experimental and theoretical bands cross
each other---appear to be quite different in the three panels of Fig.~\ref{f02}. In particular, there is no overlap between the
preferred $g_{pp}$ ranges in the upper and middle (or lower) panel, while there is only a marginal
overlap between those in the middle and lower panels. Agreement between data and theory is never reached
for all the three observables at the same time.

If one choses the $2\nu2\beta$ lifetime to fix $g_{pp}$ (as advocated in \cite{Ro07}), then two preferred ranges 
are selected, one in the positive branch (around  $g_{pp}\simeq 0.78$), and the other in the negative branch 
(around $g_{pp}\simeq 0.79$);  see the upper panel of Fig.~\ref{f02}. 
Although both ranges are phenomenologically viable, the one in the positive branch is usually adopted on
theoretical grounds \cite{Ro07}. However, for $g_{pp}\simeq0.78$,  the theoretical EC ($\beta^-$) lifetime 
turns out to be significantly smaller (larger) than the experimental value. Similar problems occur for 
$g_{pp}\simeq 0.79$ in the negative branch.

Alternatively, one might use the $\beta^{-}$ lifetime to fix $g_{pp}$ (as advocated in \cite{Su05,SuBe}). 
In this case, as evident from Fig.~\ref{f02},
one could get marginal agreement between both $\beta^-$ and EC observables around $g_{pp}\simeq 0.75$, but only at the price
of underestimating the measured $2\nu2\beta$ lifetime by a factor of $\sim\!4$. With one choice or another, it seems that
current QRPA calculations fail to reproduce all the three independent lifetimes at the same time.

The above discrepancies would become stronger by increasing the GT strength from its standard quenched value
($g_A\simeq 1$) to its bare value ($g_A\simeq 1.25$, not shown). For $g_A=1.25$, according to Eqs.~(\ref{ga1}) and (\ref{ga2}), 
the theoretical bands in Fig.~\ref{f02} would be shifted downwards by $-4\log g_A\simeq -0.4$ (upper panel) or by 
$-2\log g_A\simeq -0.2$ (middle and lower panels). The preferred ranges of $g_{pp}$ 
would then move to the right for  $2\nu2\beta$ and $\beta^-$, and to the left for
EC, thus destroying even the marginal agreement existing between $\beta^-$ and EC observables
for $g_A=1$. We conclude that, within the range $1\lesssim g_A\lesssim 1.25$, current QRPA calculations
cannot reproduce the three lifetime data (nor, to some extent, any two among them) for any value of $g_{pp}$.
These graphical results will be numerically confirmed in Sec.~\ref{SecV}.

\subsection{$^{116}$Cd data versus QRPA ($ {g_A=1}$)  \label{SecIIIB}}

\begin{figure}[t]
\begin{center}
\includegraphics[scale=1.1]{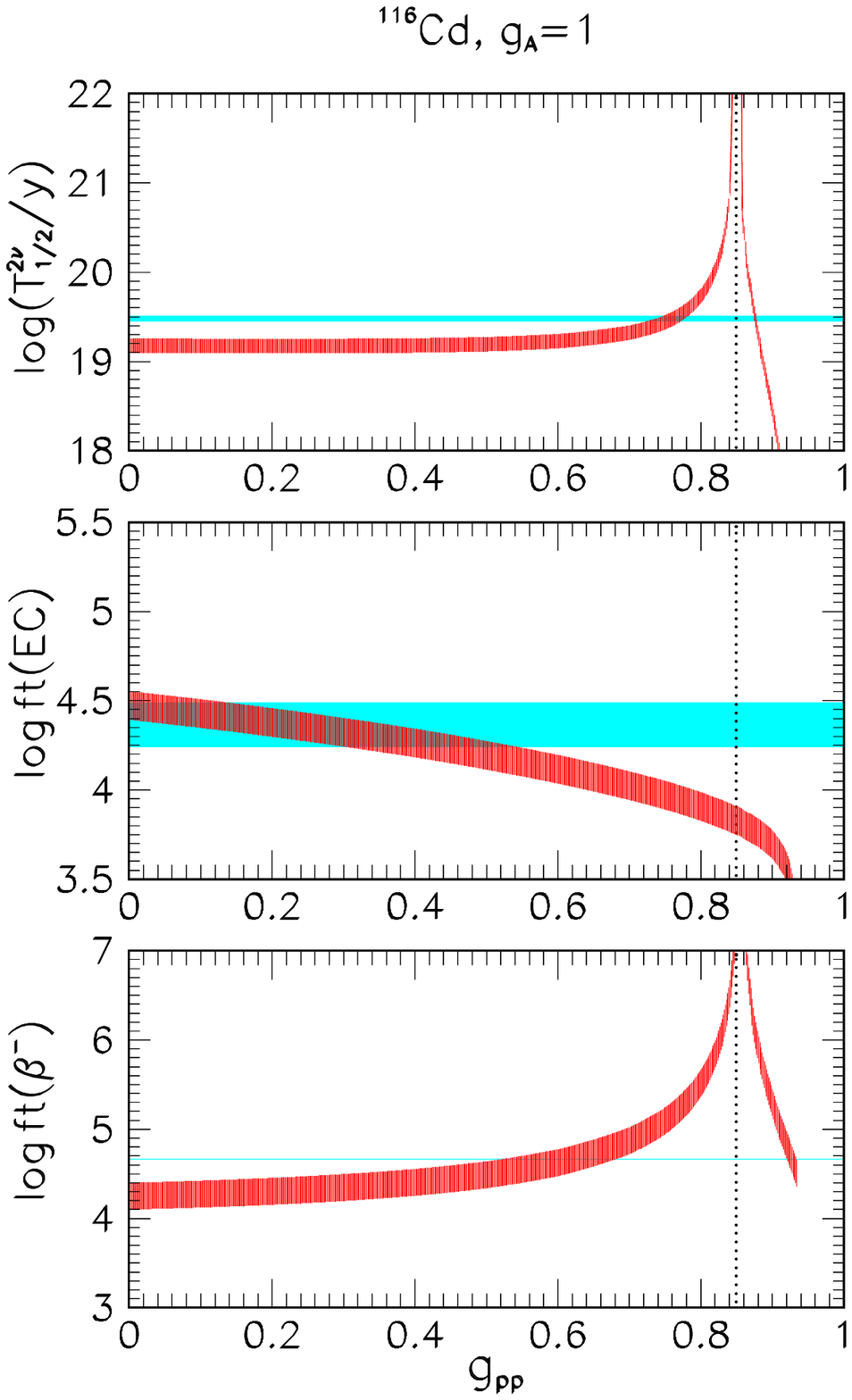}
\end{center}
\caption{ \label{f03} Lifetimes for the $2\nu2\beta$, EC, and $\beta^-$ decay in
$^{116}$Cd. Horizontal bands: experimental data. Curved bands: the QRPA results as 
a function of $g_{pp}$, for fixed $g_A=1$, in the large basis. The vertical width of 
the bands corresponds to $\pm1\sigma$ uncertainties. }
\end{figure}

Figure~\ref{f03} is analogous to Fig.~\ref{f02}, but for $^{116}$Cd. The situation is very similar to $^{100}$Mo, and the same 
qualitative considerations apply, although the preferred ranges of $g_{pp}$ are different. Also in this case,
it is not possible to reconcile the QRPA estimates with the three independent lifetime data data for any value
of $g_{pp}$, at fixed $g_A=1$. The discrepancy becomes worse for $g_{A}=1.25$ (not shown).

We remark that Figs.~\ref{f02}--\ref{f03} refer to QRPA calculations in large basis. Very similar results are 
obtained---and the same comments apply---to calculations in small basis (not shown).

\section{Data versus theory with strong quenching $ \mathbf{(g_A<1)}$ \label{SecIV}}

In the previous Section, we have shown that the QRPA fails to reproduce the three $(2\nu2\beta,\,\mathrm{EC},\,\beta^-)$ 
lifetimes in each of the two reference nuclei  ($^{100}$Mo and $^{116}$Cd), as far as $g_A$ is taken in the
usual range, $1\lesssim g_A\lesssim 1.25$. In particular, the  discrepancy becomes worse as one moves towards the
upper end of this range. Conversely, the discrepancy can be expected to
become less severe (and hopefully vanish) for $g_A<1$, corresponding to a ``strong quenching'' of the GT coupling.

Values of $g_A$ lower than unity, although rather unconventional in the 
QRPA literature, are not uncommon in NSM calculations. The NSM, being an {\em ab initio\/} approach, does 
not depend on phenomenological parameters such as $g_{pp}$, but of course retains the dependence 
on the axial coupling $g_A$, with the associated quenching uncertainties.
Although a quenched value $g_A\sim 1$ seems to roughly provide the correct normalization of the GT 
strength, strongly quenched values $g_A<1$ may occasionally be needed to bring NSM calculations in agreement
with data \cite{Ca05,Ejir}. It is fair to say that, in the NSM approach, one is not
committed to a strict range for $g_A$ (such as $1\lesssim g_A \lesssim1.25$): any value 
$g_A~\sim O(1)$ is generally accepted, if the data require so.

In both the QRPA and the NSM approach, the origin and size of the GT quenching remains in part
obscure and uncertain from a theoretical viewpoint,  and the inferred values of $g_A$ fluctuate considerably 
in different data analyses, processes, and nuclei. Even for a fixed process and nucleus, it is not excluded 
that the quenching may be energy-dependent \cite{Zi06}.  Therefore, the common practice of adopting either the 
standard quenched value $g_A\simeq 1$ or the bare value $g_A\simeq 1.25$ may be unnecessarily restrictive. 
It is perhaps more sensible to treat $g_A$ as a free parameter of order
unity, whose precise value needs to be constrained by the data themselves,
rather than pre-assigned by theory---just as one does for $g_{pp}$.
In the following, we thus adopt a purely phenomenological viewpoint, and show that specific choices of $g_A$
below unity (which will be more precisely derived in Sec.~\ref{SecV}) can bring QRPA calculations in agreement
with all the three lifetime data in each of two reference nuclei.

\subsection{$^{100}$Mo data versus QRPA ($ {g_A=0.74}$) \label{SecIVA}}

\begin{figure}[t]
\begin{center}
\includegraphics[scale=1.1]{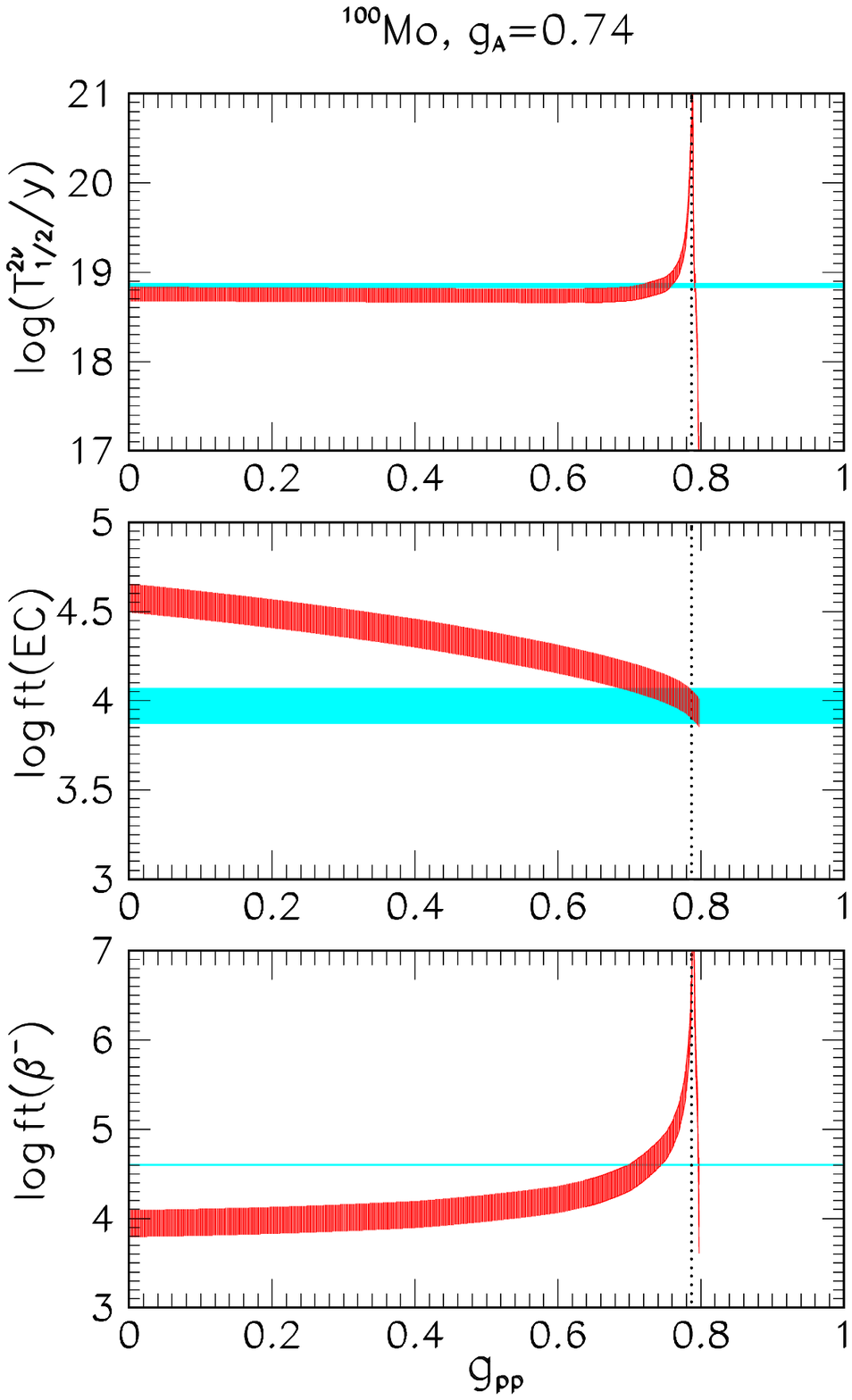}
\end{center}
\caption{ \label{f05} As in Fig.~\protect\ref{f02}, but for $g_A=0.74$.
Note the overall agreement of the QRPA results with the data for $g_{pp}\sim 0.73$. }
\end{figure}

Figure~\ref{f05} is analogous to Fig.~\ref{f02}, but for the strongly quenched value $g_A=0.74$. 
We anticipate that this value provides the
best overall agreement of QRPA calculations (curved bands) with the data (horizontal bands). 
Around $g_{pp}\simeq 0.73$, all bands cross each other in the three panels. No such
common crossing occurs in the negative branch, as also confirmed by numerical explorations. Besides
selecting the positive branch, the data appear to
prefer a particle-particle strength ($g_{pp}\simeq 0.73$) sufficiently far from the
the critical and collapse values, where the QRPA estimates become less reliable. 
The $\beta^-$ theoretical band in the lower panel is rather steep around $g_{pp}\simeq 0.73$, and can
thus provide, together with the experimental datum, both an upper and a lower bound to $g_{pp}$;
the upper (lower) bound can also be enforced by the $2\nu2\beta$ (EC) observable, as evident in the
upper (middle) panel. In perspective, a reduction of the EC error in the middle panel would be beneficial
to better probe this strong-quenching scenario.

\subsection{$^{116}$Cd data versus QRPA ($ {g_A=0.84}$) \label{SecIVB}}

\begin{figure}[t]
\begin{center}
\includegraphics[scale=1.1]{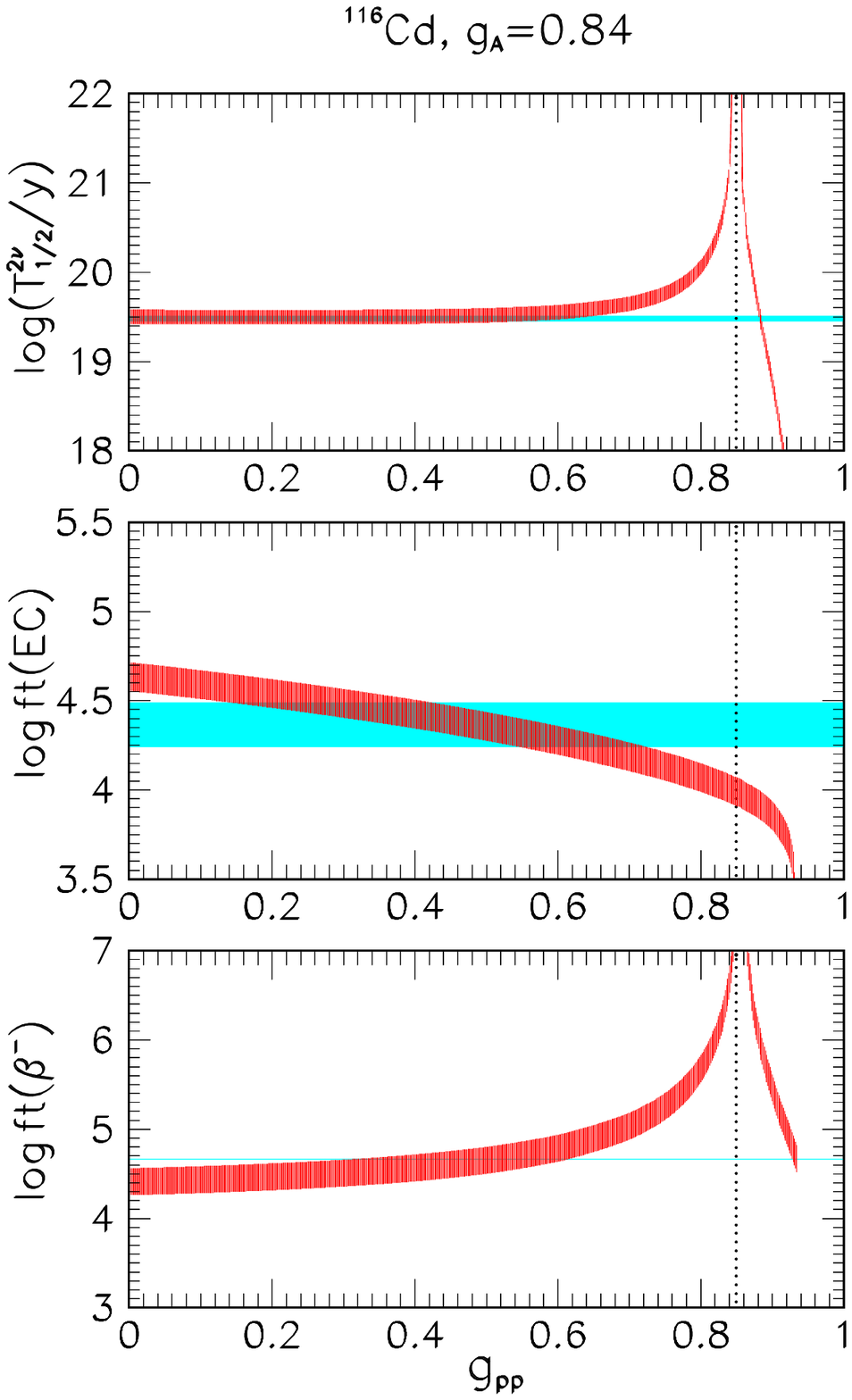}
\end{center}
\caption{ \label{f06} As in Fig.~\protect\ref{f03}, but for $g_A=0.84$. 
Note the overall agreement of the QRPA results with the data for $g_{pp}\sim 0.5$. }
\end{figure}

Figure~\ref{f06} is analogous to Fig.~\ref{f03}, but for the strongly quenched value $g_A=0.84$. A good overall 
agreement between theory and data is reached in a broad range $g_{pp}\simeq 0.4$--0.6. It is interesting to 
note that this range could be significantly restricted if the experimental errors of the EC datum \cite{Bhat} in 
the middle panel were reduced by a factor of two or more. Also in this case, the
data unambiguously select the positive branch, and keep $g_{pp}$ far from the critical and collapse
points.

We remark that Figs.~\ref{f05}--\ref{f06} refer to QRPA calculations in large basis. Very similar results 
are obtained---and the same comments apply---to calculations in small basis (not shown).

\subsection{Discussion \label{SecIVD}}

Strong quenching $(g_A<1)$ appears to provide a phenomenological solution to the well-known overall discrepancy 
between  QRPA results and lifetime data. This solution is nontrivial because: 
(1) two free parameters enable to reproduce very well, 
within $1\sigma$ uncertainties, three independent data (in each of two different nuclei); 
(2) the positive branch $(M^{2\nu}>0)$, which is favored by theoretical arguments,
is unambiguously selected by the data; 
(3) the preferred values of $g_{pp}$ are far enough from the critical and collapse values. 
Such data-driven features seem to be more than accidental facts, and suggest that $g_A<1$ 
might be a realistic option within the QRPA. More accurate lifetime data (especially for EC and, to some extent, for
$2\nu2\beta$ decay),  as well as further charge-exchange reaction data
(not considered in this work) should provide additional probes of the strong quenching solution.

This solution is admittedly unconventional in the context of QRPA, where $g_A$ has been customarily 
taken within the range $1\lesssim g_A\lesssim 1.25$. 
It may be that 
strong quenching is associated to other effects, 
whose degrees of freedom might be traded for milder variations of $g_A$. 
However, if new free parameters are added to $g_{pp}$ and $g_A$, the data set must also be enlarged to provide 
meaningful and nontrivial constraints---not much would be learned, in general, by fitting $N$ data
with $\geq N$ parameters.

For the sake of simplicity, in this work we do not explore more elaborate scenarios with 
additional data and further QRPA degrees of freedom. We just take for granted the indication in 
favor of $g_A<1$, and perform quantitative fits to three
selected data (the $2\nu2\beta$, EC, and $\beta^-$ lifetimes) via two parameters $(g_{pp},~g_A)$. We shall thus
obtain an overconstrained parameter space, used for subsequent
$0\nu2\beta$ calculations in Sec.~\ref{SecVI}. Despite the above caveats, 
this approach represents a step forward
with respect to previous attempts, which aimed at reducing the $0\nu2\beta$ model 
uncertainties in QRPA by fitting a single datum (either $2\nu2\beta$ or $\beta^-$) through a single parameter ($g_{pp}$) 
at fixed $g_A$.

\section{Overconstraining the ($ {g_{pp},\,g_A}$) parameters  \label{SecV}}

We perform a least-square fit to the three data $x_1=\log(T^{2\nu}_{1/2}/\mathrm{y})$, $x_2=\log ft(\mathrm{EC})$, 
and $x_3=\log ft(\beta^-)$
in terms of the two free parameters $(g_{pp},\, g_A)$. The $\chi^2$ function to be minimized is defined as
\begin{equation}
\chi^2 (g_{pp},\,g_A) =\sum_{i=1}^3 
\frac{\left[x_i^\mathrm{exp}-x_i^\mathrm{th}(g_{pp},\,g_A)\right]^2}{(\sigma^i_\mathrm{exp})^2+(\sigma^i_\mathrm{th})^2}\ .
\label{chi2}
\end{equation}
where all the ingredients have been defined in the previous Sections. Asymmetric experimental errors 
(see Tab.~\ref{tab:inputdata}) are properly included by choosing either the upper or lower error, according to the sign 
of the difference $x_i^\mathrm{th}-x_i^\mathrm{exp}$. The minimum search is performed by numerical scan
over a dense grid in the $(g_{pp},\,g_A)$ rectangle $[0,\,1]\otimes[0,\,1.25]$. 
Given three data and two parameters, one expects $\chi^2_{\min}\sim O(1)$ for a proper fit. The expansion around the 
best-fit values of $(g_{pp},\, g_A)$ at $\Delta \chi^2=\chi^2-\chi^2_{\min} = n^2$   provides then the $n$-$\sigma$ contours
for such parameters \cite{PDGr}. In the following, we show the main results both in graphical and tabular form.

Figure~\ref{f08} shows the results of the $(g_{pp},~g_A)$ fit in large basis. In each of the two panels
(corresponding, from top to bottom, to $^{100}$Mo and $^{116}$Cd) a dot marks the best-fit point, 
surrounded by the 1, 2 and $3\sigma$ contours. Vertical dotted lines
separate the positive and negative branches of $M^{2\nu}$.
In both panels, the allowed regions are fully contained in the positive branch, thus confirming quantitatively
the theoretical arguments in favor of $M^{2\nu}>0$ \cite{Ro07}. The best-fit points
are safely far from extremal values of $g_{pp}$ (0 and $\sim\!g_{pp}^*$), but the allowed regions
may extend towards one of them. In particular, the allowed range of $g_{pp}$ is somewhat squeezed
towards the critical value for $^{100}$Mo, while it extends towards zero for $^{116}$Cd at $3\sigma$. 
More accurate experimental data (especially from EC and, to some extent, from $2\nu2\beta$ decay) would
be useful to shrink such ranges, as discussed in Sec.~\ref{SecIV}, and might thus prevent the occurrence
of nearly extremal values of $g_{pp}$. Concerning $g_A$, strong quenching $(g_A<1)$ is definitely preferred at 
$>3\sigma$ in both cases.

We emphasize that ``overconstraining the $(g_{pp},~g_A)$ parameters'' is equivalent to
state that, in each of the $^{100}$Mo and $^{116}$Cd reference nuclei, our scenario 
makes one prediction which is experimentally verified. Figures~\ref{f09} and \ref{f099} illustrate this statement
via the $1\sigma$ bands individually allowed by $\beta^-$, EC and $2\nu2\beta$ data for $^{100}$Mo and $^{116}$Cd,
as obtained by a breakdown of the three contributions in Eq.~(\ref{chi2}). Any two
bands can be used to constrain $(g_{pp},~g_A)$ in a closed region (the ``prediction''), which is then 
crossed by the third independent band (the ``experimental verification'').

The numerical results of the global $(g_{pp},~g_A)$ fit in large basis
are summarized in Table~\ref{tab:fitlarge}. The fit quality is very good in all cases ($\chi^2_{\min}\lesssim 1$)
and the best-fit values for the three lifetimes are in striking agreement with the corresponding data
in Tab.~\ref{tab:inputdata}, which are repeated for convenience  in Table~\ref{tab:fitlarge} (in square brackets).
The best-fit values and $\pm n\sigma$ ranges ($n=1,\,2,\,3$) for $g_{pp}$ and
$g_A$ are also reported. (The $g_A$ values adopted in Figs.~\ref{f05} and \ref{f06} are just taken from Table~\ref{tab:fitlarge}.)


\Table{\label{tab:fitlarge}  Results of the $(g_{pp},\,g_A)$ fit for two different (mother) nuclei, with 
QRPA calculations performed in large basis. Column 2: minimum $\chi^2$. Columns 3--5:
theoretical lifetimes for $2\nu2\beta$, EC and $\beta^{-}$ decay at best fit, to be compared with the experimental 
data in  Tab.~\protect\ref{tab:inputdata} which are repeated here in square brackets. 
Columns 6--9: value of $g_{pp}$ at best fit, and allowed ranges at 1, 2 and 3$\sigma$. 
Columns 10--13: value of $g_{A}$ at best fit, and allowed ranges at 1, 2 and 3$\sigma$.}
\br
Nuclei & $\chi^2_{\min}$ 
& $\log(T^{2\nu}_{1/2}/\mathrm{y})$ & $\log ft(\mathrm{EC})$  & $\log ft(\beta^-)$
& $g_{pp}$ \ \ \ $\pm1\sigma$\ \ \ $\pm2\sigma$\ \ \ $\pm3\sigma$ 
&  $g_{A}$ \ \ \ \  $\pm1\sigma$  \ \  $\pm2\sigma$  \ \  $\pm3\sigma$  \\[1mm]
\mr
\noalign{\vspace*{1mm}}
$^{100}$Mo& 1.26 & 18.82 [18.85]& 4.09 [3.96]& 4.66 [4.60]& 0.733 $^{+0.020}_{-0.020}$ $^{+0.031}_{-0.063}$ $^{+0.039}_{-0.126}$ & 0.741 $^{+0.046}_{-0.037}$ $^{+0.120}_{-0.074}$ $^{+0.176}_{-0.107}$ \\[2mm]
$^{116}$Cd& 0.12 & 19.49 [19.48]& 4.35 [4.39]& 4.63 [4.66]& 0.493 $^{+0.106}_{-0.149}$ $^{+0.173}_{-0.358}$ $^{+0.224}_{-0.493}$ & 0.843 $^{+0.042}_{-0.037}$ $^{+0.088}_{-0.075}$ $^{+0.149}_{-0.106}$\\ 
\br
\endTable

We have repeated the analysis in small basis, with similar results. The graphical results are omitted, while the numerical 
ones are reported in Table~\ref{tab:fitsmall}. The quality of the fit is very good also in this case.
The allowed ranges for $g_{pp}$ and $g_A$ in small basis (Tab.~\ref{tab:fitsmall}) are somewhat different from those 
in large basis  (Table~\ref{tab:fitlarge}), but with similar features. In particular, the allowed $g_{pp}$ range is
in the positive branch, and the general trend in favor of
$g_A<1$ is confirmed. We conclude that the main results obtained so far 
do not change qualitatively with the size of the basis.

\Table{\label{tab:fitsmall} As in Table~\protect\ref{tab:fitlarge}, but in small basis.}
\br
Nuclei & $\chi^2_{\min}$ 
& $\log(T^{2\nu}_{1/2}/\mathrm{y})$ & $\log ft(\mathrm{EC})$      & $\log ft(\beta^-)$
& $g_{pp}$ \ \ \ $\pm1\sigma$\ \ \ $\pm2\sigma$\ \ \ $\pm3\sigma$ 
&  $g_{A}$ \ \ \ \  $\pm1\sigma$  \ \  $\pm2\sigma$  \ \  $\pm3\sigma$ \\[1mm]
\mr
\noalign{\vspace*{1mm}}
$^{100}$Mo& 1.11 & 18.82 [18.85] & 4.08 [3.96] & 4.67 [4.60] & 0.862 $^{+0.024}_{-0.035}$  $^{+0.043}_{-0.094}$  $^{+0.055}_{-0.181}$ & 0.745 $^{+0.042}_{-0.037}$ $^{+0.098}_{-0.074}$ $^{+0.172}_{-0.111}$ \\[2mm]
$^{116}$Cd& 0.03 & 19.49 [19.48] & 4.37 [4.39] & 4.65 [4.66] & 0.540 $^{+0.130}_{-0.165}$ $^{+0.220}_{-0.385}$ $^{+0.283}_{-0.538}$ & 0.815 $^{+0.042}_{-0.033}$ $^{+0.084}_{-0.070}$ $^{+0.139}_{-0.102}$ \\
\br
\endTable
\medskip
\medskip
\medskip
\medskip

\section{Implications for $0\nu2\beta$ decay \label{SecVI}}

In the previous Section we have obtained allowed regions in the parameter space $(g_{pp},\,g_A)$. 
In this Section we study how such regions affect the QRPA calculation of $0\nu2\beta$ decay, after
recalling some basic features of this process.

The $(2\nu2\beta,\,\mathrm{EC},\,\beta^-)$ processes that we have considered
so far occur only via GT transitions through $1^+$ intermediate states. The leading contribution $M^{0\nu}_\mathrm{GT}$ 
to the amplitude of the neutrinoless double beta decay also comes from the GT-type transitions which, 
however, proceed through  intermediate states of all, but $0^+$, multipolarities. In addition,
there are Fermi $M^{0\nu}_\mathrm{F}$ and (small) tensor $M^{0\nu}_\mathrm{T}$ contributions 
to the $0\nu2\beta$ matrix element,
\begin{equation}
\label{matr}
M^{0\nu}(g_{pp},g_A)= M^{0\nu}_\mathrm{GT}(g_{pp})+
M^{0\nu}_\mathrm{T}(g_{pp})-\displaystyle\frac{M^{0\nu}_\mathrm{F}(g_{pp})}{g^2_{A}}\ ,
\end{equation}
where the dependence on $g_{pp}$ and $g_A$ is made explicit.

\begin{figure}[t]
\begin{center}
\includegraphics[scale=1.1]{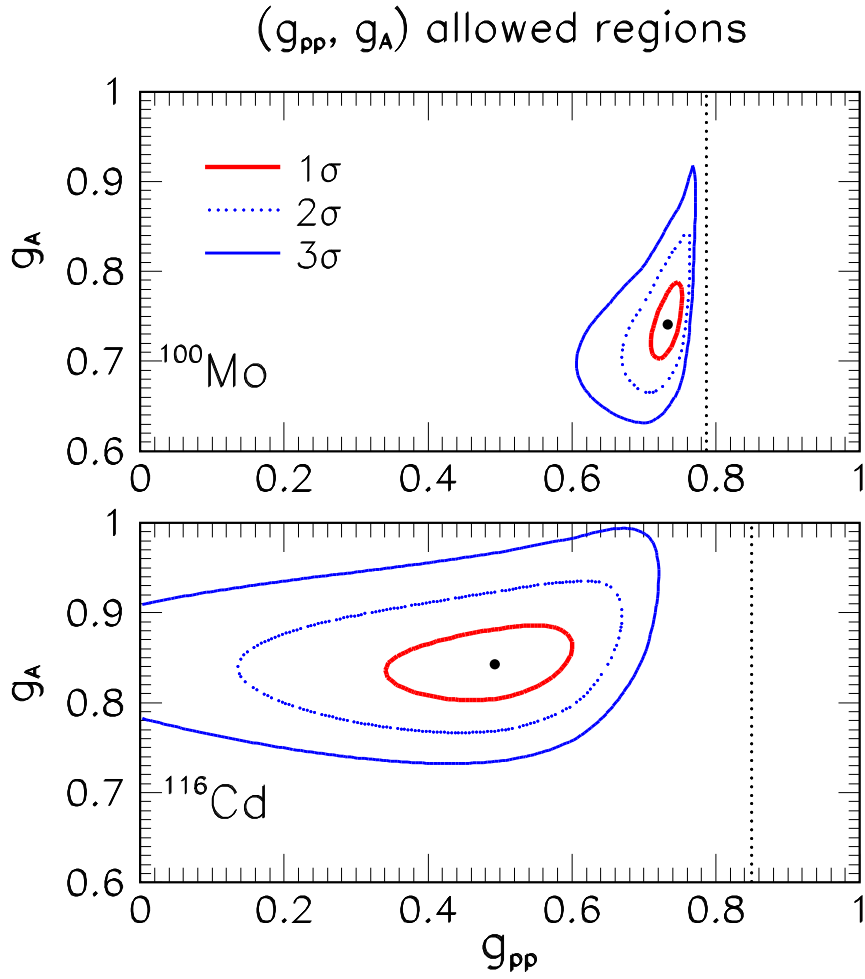}
\end{center}
\caption{ \label{f08} Regions allowed at $n$-$\sigma$ in the $(g_{pp},\,g_A)$ plane 
from a QRPA fit to  the $2\nu2\beta$, EC, and $\beta^-$ data, in each of the two nuclei
$^{100}$Mo and $^{116}$Cd. The QRPA calculations refer to the large basis.}
\end{figure}

\begin{figure}[t]
\begin{center}
\includegraphics[scale=1.1]{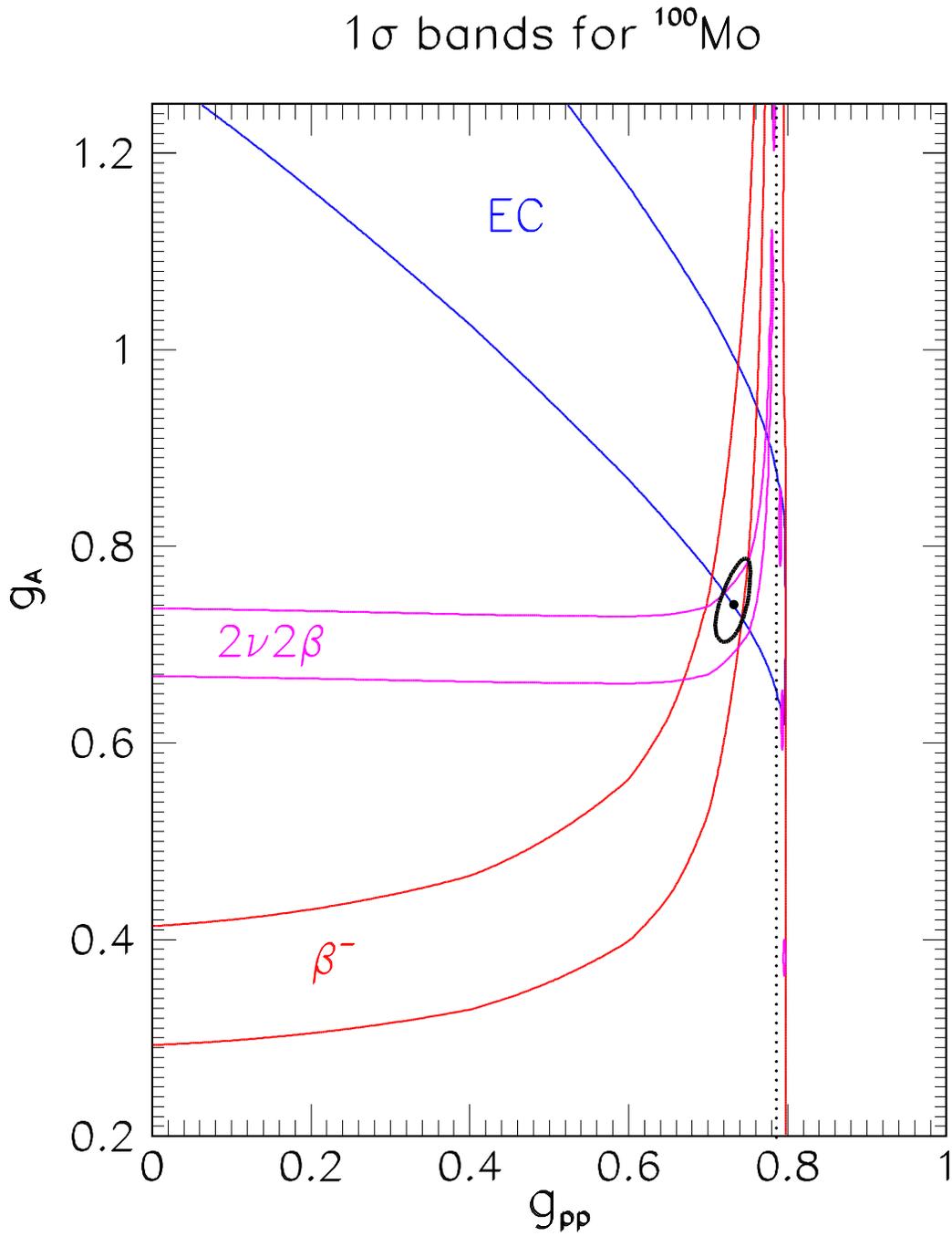}
\end{center}
\caption{ \label{f09} Breakdown of individual constraints in the $(g_{pp},\,g_A)$ plane
for $^{100}$Mo. The slanted bands corresponds to the regions allowed at $1\sigma$ level
(including experimental and theoretical errors) by $\beta^-$, EC, and $2\nu2\beta$ data.
Their combination (thick ellipse) coincides with the $1\sigma$ contour in the upper plot
of Fig.~\protect\ref{f08}.}
\end{figure}

\begin{figure}[t]
\begin{center}
\includegraphics[scale=1.1]{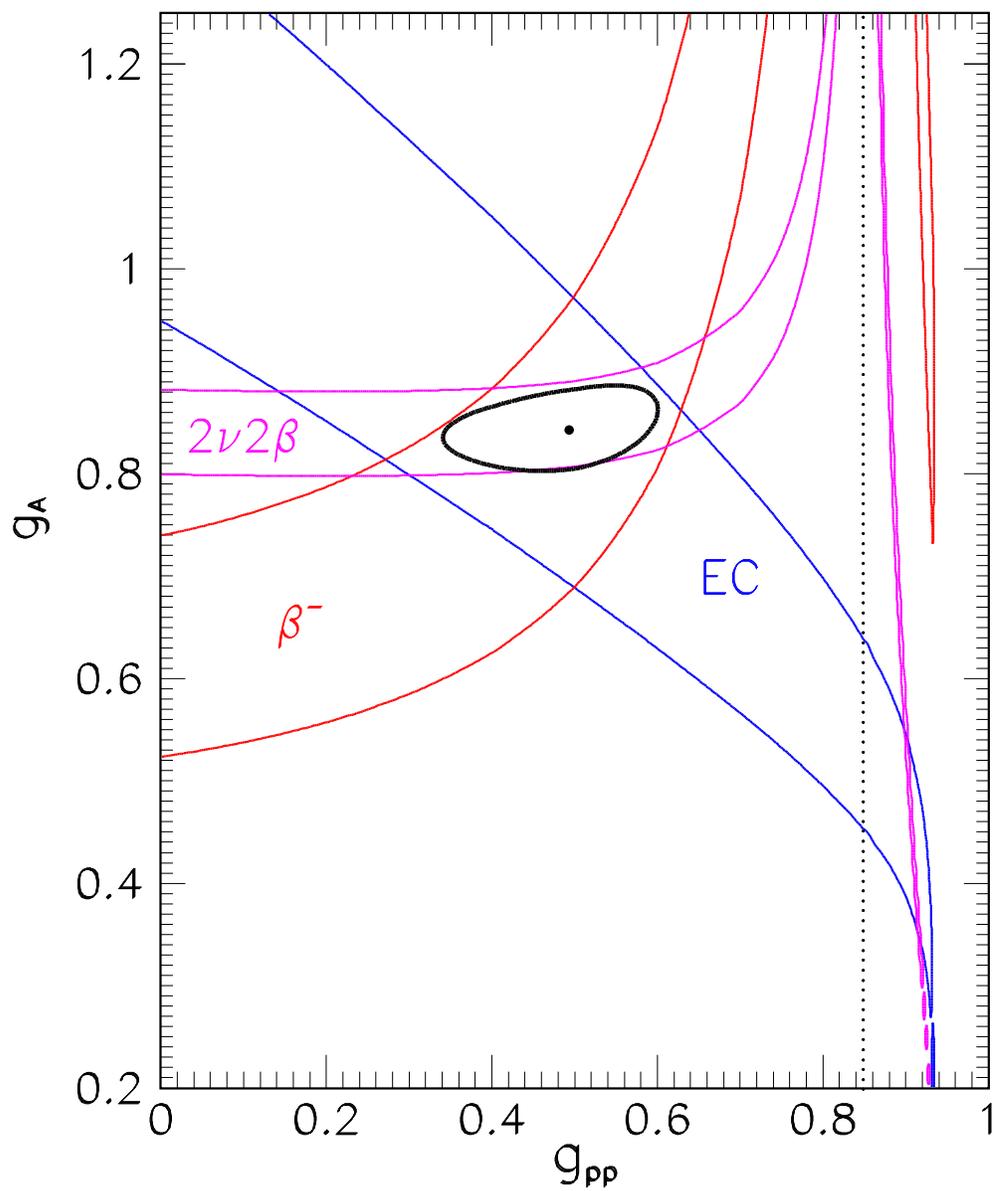}
\end{center}
\caption{ \label{f099} As in Fig.~\protect\ref{f09}, but for $^{116}$Cd. The thick ellipse coincides with the $1\sigma$ contour in the lower plot of Fig.~\protect\ref{f08}.}
\end{figure}

\begin{figure}[t]
\begin{center}
\includegraphics[scale=1.1]{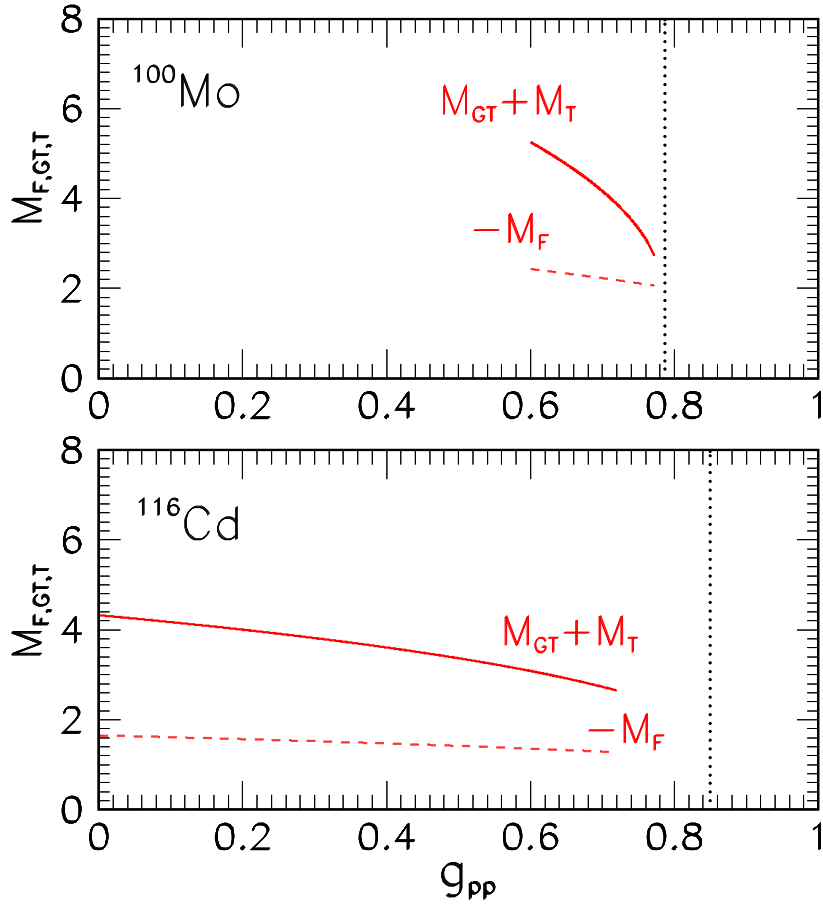}
\end{center}
\caption{ \label{f10} $0\nu2\beta$ matrix element components  
$M^{0\nu}_\mathrm{GT}+M^{0\nu}_\mathrm{T}$ (solid) and $-M^{0\nu}_\mathrm{F}$ (dashed),
as a function of the $g_{pp}$ parameter in its $3\sigma$ allowed range (see  Fig.~\protect\ref{f08}).
The QRPA calculations refer to the default case (the large basis with the Jastrow-like short range correlations).  }
\end{figure}

Figure~\ref{f10} shows the relevant components of the $0\nu2\beta$ matrix elements as a 
function of $g_{pp}$ in large basis, and including short range correlations, which will be shortly discussed below. 
Since the QRPA calculation is computer-intensive, $g_{pp}$ is varied only within the relevant $\pm3\sigma$ range 
shown in Fig.~\ref{f08}. Note that the leading component shows significant
variations with $g_{pp}$, so that any constraint on this parameter (such as those
derived in the previous Section) helps to reduce the spread of
QRPA estimates of $0\nu2\beta$ decay.
Results qualitatively similar to Fig.~\ref{f10} 
are obtained for small basis, or without short range correlations (not shown).

Given the QRPA results in Fig.~\ref{f10}, the $0\nu2\beta$ matrix element can be computed for any relevant
value of $g_A$ and $g_{pp}$ through Eq.~(\ref{matr}).  In order to make contact with the notation in Ref.~\cite{Ro07}, 
we shall actually rescale the matrix element as
\begin{equation}
M'^{0\nu}= M^{0\nu}\left(\frac{g_{A}}{1.25}\right)^2\ .
\end{equation}
The $0\nu2\beta$ lifetime  reads then
\begin{equation}
T^{0\nu}(g_{pp},g_A)=\frac{t^{0\nu}_{1/2}}{|M'^{0\nu}|^2}\ ,
\end{equation}
where the proportionality factor $t^{0\nu}_{1/2}=\left(m_{\beta\beta}^2 G^{0\nu}/1.25^2\right)^{-1}$ [y] is 
numerically given by
\begin{equation}
t^{0\nu}_{1/2}=\left\{ 
\begin{array}{ll}
1.83\times 10^{27} & (^{100}\mathrm{Mo})\ ,\\
1.68\times 10^{27} & (^{116}\mathrm{Cd})\ ,
\end{array}
\right.\ 
\end{equation}
for a reference Majorana mass $m_{\beta\beta}=50$ meV. For different values of $m_{\beta\beta}$, one just 
rescales $t^{0\nu}_{1/2}\propto m_{\beta\beta}^{-2}$.

For any given value of $(g_{pp},g_{A})$, calculations of $M'^{0\nu}$ are affected 
not only by the size of the basis (either large or small), but also by uncertainties which are
peculiar of the $0\nu2\beta$ process, namely, those related to the important issue of short range correlations 
(s.r.c.). These correlations account for the well-known fact that the nucleon-nucleon interaction becomes 
strongly repulsive at small internucleon distances. This in turn must lead to
strong suppression of the relative-motion wave function at small distances (s.r.c.\ effects).
Short range correlations are explicitly included neither within the QRPA nor within the NSM. They are instead
introduced {\it ad hoc} directly into the neutrino potential via a multiplicative factor (the square of a correlation function). 
One of the most popular is the Jastrow-like correlation function~\cite{MS1976} which has been used 
in the previous calculations~\cite{Ro03,Ro07} and is also used in this work. 
We shall thus present results in four cases, 
corresponding to either large or small basis, with or without the Jastrow-like s.r.c.\ effects.

In each of the four cases, the effect of the $(g_{pp},\,g_A)$ uncertainties on $M'^{0\nu}$ is estimated by 
marginalization \cite{PDGr}, taking into account the fact that the same fixed value for the matrix
element may be realized by different (``degenerate'') couples of values $(g_{pp},~g_A)$.
More precisely, given the function $\chi^2(g_{pp},\,g_A)$ defined in the previous
Section, and for a fixed value $\tilde M'^{0\nu}$, we define a marginalized $\chi^2$ function,
\begin{equation}
\chi^2(\tilde M'^{0\nu})=\min_{\tilde g_{pp},\tilde g_A} \chi^2(\tilde g_{pp},\tilde g_A)\ ,
\end{equation}
over the degenerate set of $(\tilde g_{pp},\tilde g_A)$ obeying
\begin{equation}
 M'^{0_\nu}(\tilde g_{pp},\tilde g_A)=  \tilde M'^{0\nu}\ .
\end{equation}
The minimization of $\chi^2(\tilde M'^{0\nu})$, and the expansion around the minimum at $\Delta\chi^2=n^2$, provide
the correct best-fit values and $n\sigma$ ranges for $M'^{0\nu}$, respectively. 
Since we are interested in $n\leq 3$, we perform a numerical marginalization
over a dense, rectangular grid covering only the $\pm3\sigma$ ranges of $(g_{pp},~g_A)$.

Tables~\ref{tab:withsrc} and \ref{tab:nosrc}
provide an overview of the derived ranges for $M'^{0\nu}$ at 1, 2 and $3\sigma$ (in large and small basis),
with and without the effect of s.r.c., respectively. We also report the corresponding ranges
for the measurable (log) lifetime $T^{0\nu}_{1/2}$, at the reference value $m_{\beta}=50$~meV.
Note the $\pm n\sigma$ ranges are generally asymmetric and do not scale
linearly, in part as a consequence of the original one-sided $g_{pp}$ limits at either 0 or $\sim\!g_{pp}^*$ (see Fig.~\ref{f08}). 
By comparing the results in Tables~\ref{tab:withsrc} and \ref{tab:nosrc}, it appears that
the basis size is not the major source of systematic uncertainties. Conversely, the inclusion or exclusion of s.r.c.\ effects
always induce changes  $>1\sigma$. 

\begin{figure}[t]
\begin{center}
\includegraphics[scale=1.1]{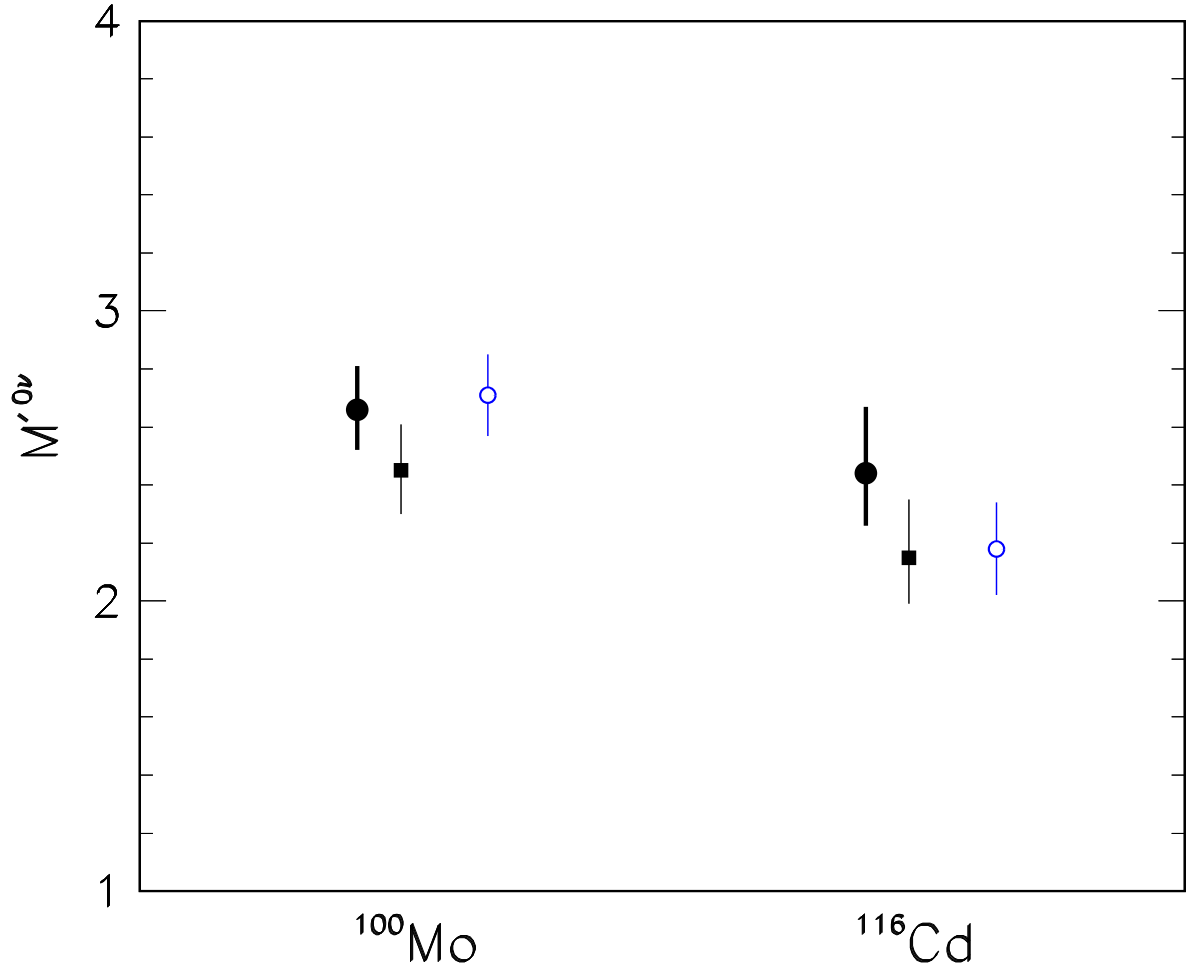}
\end{center}
\caption{ \label{f11} Overview of $0\nu2\beta$ matrix elements $M'^{0\nu}$, together with their
$\pm1\sigma$ estimated errors. For each nucleus, three QRPA cases are shown. 
From the left to right, the first two cases correspond to the results
of this work in the large basis (black circle, with thick error bars) and in
the small basis (black square). The error bars for these two cases encompass the uncertainties
in both parameters $(g_{pp},\,g_A)$ from the fit to $(2\nu2\beta,\,\mathrm{EC},\,\beta^-)$ data. 
The third case (white circle) refers to the previous results of
Ref.~\protect\cite{Ro07}, as obtained for the fixed value $g_A=1$ (with $g_{pp}$ adjusted to $2\nu2\beta$ 
data).  All cases include the effects of the Jastrow-like s.r.c.}
\end{figure}

Figure~\ref{f11} shows an overview of QRPA results for the nuclear matrix elements (including s.r.c.\ effects)
in three different cases  for each nucleus. From left to right, the first two cases
correspond to the $1\sigma$ ranges from Table~\ref{tab:withsrc}, in large and small basis, respectively. 
The third case correspond to the results previously obtained in \cite{Ro07} for $g_A=1$ (with correspondingly  smaller error bars, due to the fixed $g_A$ value). 
Remarkably, such results for $M'^{0\nu}$ \cite{Ro07} differ by $\lesssim 12\%$ from those obtained in this work, 
in spite of a marked difference in the central values of $g_A$ and $g_{pp}$. 

Summarizing, in each of the two nuclei examined it is possible: 
($i$) to fit very well three data ($2\nu2\beta$, EC, $\beta^-$) with two parameters ($g_{pp},g_A$), provided that $g_A<1$; 
($ii$) to exclude the negative branch $M^{2\nu}<0$; and 
($iii$) to derive robust ranges for $0\nu2\beta$ observables. 
There remains a relative large uncertainty on the $0\nu2\beta$ matrix element, associated with the
size of short range correlation effects. 
Unfortunately, s.r.c.\ effects are peculiar of $0\nu2\beta$ decay and are not constrained at all by the 
($2\nu2\beta$, EC, $\beta^-$) data considered in this work.

\Table{\label{tab:withsrc} QRPA estimates of the $0\nu2\beta$ matrix elements 
$M'^{0\nu}$, including the effect of short range correlations. The central value and the
allowed ranges of $M'^{0\nu}$ are derived, respectively, from the best-fit values and allowed ranges of $(g_{pp},g_A)$.
The estimates refer to both large basis (l.b.) and small basis (s.b.). We also report the corresponding (logarithmic) ranges
for the $0\nu2\beta$ lifetime $T^{0\nu}_{1/2}$ (in years), assuming a reference value $m_{\beta\beta}=50$~meV.}
\br
Nucleus & 
$M'^{0\nu}$& $\pm1\sigma$& $\pm2\sigma$& $\pm3\sigma$ & 
$\log(T^{0\nu}_{1/2})$ & $\pm1\sigma$& $\pm2\sigma$& $\pm3\sigma$ 
\\[1mm]
\noalign{\vspace*{1mm}}
\mr
Large basis&&&&&&&&\\[2mm]
$^{100}$Mo & 2.66 &$^{+0.15}_{-0.14}$&$^{+0.33}_{-0.25}$&$^{+0.61}_{-0.35}$& 
 26.411 &$^{+0.046}_{-0.046}$&$^{+0.088}_{-0.100}$&$^{+0.124}_{-0.180}$\\[1mm]
$^{116}$Cd & 2.44 &$^{+0.23}_{-0.18}$&$^{+0.53}_{-0.32}$&$^{+0.90}_{-0.44}$& 
 26.448 &$^{+0.065}_{-0.079}$&$^{+0.123}_{-0.169}$&$^{+0.174}_{-0.272}$\\[2mm]
\mr
Small basis&&&&&&&&\\[2mm]
$^{100}$Mo & 2.45 &$^{+0.16}_{-0.15}$&$^{+0.35}_{-0.25}$&$^{+0.65}_{-0.34}$&
 26.485 &$^{+0.055}_{-0.055}$&$^{+0.095}_{-0.116}$&$^{+0.132}_{-0.204}$\\[1mm]
$^{116}$Cd & 2.15 &$^{+0.20}_{-0.16}$&$^{+0.46}_{-0.29}$&$^{+0.78}_{-0.41}$&
26.561 &$^{+0.067}_{-0.079}$&$^{+0.127}_{-0.169}$&$^{+0.181}_{-0.271}$\\[2mm]
\br
\endTable

\Table{\label{tab:nosrc} As in Tab.~\ref{tab:withsrc}, but without short range correlations. }
\br
Nucleus & 
$M'^{0\nu}$& $\pm1\sigma$& $\pm2\sigma$& $\pm3\sigma$ & 
$\log(T^{0\nu}_{1/2})$ & $\pm1\sigma$& $\pm2\sigma$& $\pm3\sigma$ 
\\[1mm]
\noalign{\vspace*{1mm}}
\mr
Large basis&&&&&&&&\\[2mm]
$^{100}$Mo& 3.27 &$^{+0.16}_{-0.15}$&$^{+0.34}_{-0.29}$&$^{+0.63}_{-0.42}$&
 26.233 &$^{+0.041}_{-0.043}$&$^{+0.080}_{-0.087}$&$^{+0.117}_{-0.155}$\\[2mm]
$^{116}$Cd& 2.84 &$^{+0.25}_{-0.19}$&$^{+0.57}_{-0.35}$&$^{+0.98}_{-0.49}$&
 26.317 &$^{+0.061}_{-0.073}$&$^{+0.116}_{-0.159}$&$^{+0.165}_{-0.256}$\\[2mm]         
\mr
Small basis&&&&&&&&\\[2mm]
$^{100}$Mo& 2.97 &$^{+0.17}_{-0.16}$&$^{+0.36}_{-0.29}$&$^{+0.67}_{-0.40}$&
 26.318 &$^{+0.048}_{-0.049}$&$^{+0.089}_{-0.100}$&$^{+0.125}_{-0.178}$\\[2mm]
$^{116}$Cd& 2.47 &$^{+0.22}_{-0.17}$&$^{+0.49}_{-0.32}$&$^{+0.84}_{-0.44}$&
 26.440 &$^{+0.063}_{-0.075}$&$^{+0.119}_{-0.159}$&$^{+0.171}_{-0.255}$
\\[2mm]         
\br
\endTable

\section{Conclusions and Perspectives \label{SecVII}}

It was shown in~\cite{Ro07} that, by fitting $g_{pp}$ in order to reproduce in calculations 
the corresponding experimental $2\nu2\beta$ decay lifetimes, 
the sensitivity of calculated $0\nu2\beta$ matrix elements to other ingredients of the QRPA, such as 
the basis size, can be successfully removed. Also, it was shown that the sensitivities of the results to 
$g_A$ gets much milder than one could naively expect. There are also different proposals for fixing 
$g_{pp}$, for instance, by reproducing the single beta decay observables as advocated in~\cite{Ci05,Su05}.
By fitting $g_{pp}$ to reproduce the $\beta^-$ lifetimes of the ground states of the intermediate nucleus 
one gets the results which are similar to the ones obtained in~\cite{Ro07}, but the EC or $2\nu2\beta$ lifetimes 
are not reproduced. In this paper we have tried to reconcile all these data 
(available for the two nuclei $^{100}$Mo and $^{116}$Cd) by letting $g_A$ to be a free 
parameter of the model. In each nucleus, we have then found systematic indications in favor of strong quenching $(g_A<1)$, 
and we have been able to overconstrain two parameters $(g_{pp},\,g_A)$ with three lifetime data ($2\nu2\beta$, EC, $\beta^-$), as well as to fix the sign of $M^{2\nu}$ ($>0$).

The quenched values of $g_A$ for 
$A=100$ and $A=116$
nuclear systems obtained in this work ($g_A\simeq 0.74$ and $g_A \simeq 0.84$, respectively), although a bit unusual, 
are not much below the typical range $g_A \simeq 0.9$--1.0 
(corresponding to the quenching factor $q=g_A/1.25 \approx 0.7$--0.8) 
used within the NSM for lighter nuclei~\cite{Ca05}. 
Even stronger quenching $q\sim 0.5$ (corresponding in our notation to $g_A \sim 0.6$) has been called for 
in shell model calculations~\cite{Sk93,Br94,Juo05} of the Gamow-Teller strength for nuclei in the region of $A \sim 100$, to which the systems considered in the present work are close.

The physical origin of the quenching of $g_A$ has been discussed in the past. 
One explanation \cite{Bor81} assigns this effect to the $\Delta$-isobar admixture 
in the nuclear wave function. Another---more generally accepted---explanation \cite{Ber82} 
assigns the quenching to the shift of the Gamow-Teller strength to higher
excitation energies due to the short range tensor correlations. In light nuclei the 
quenching found in M1 transitions reduces $g_A$ from its bare value ($\sim\!1.25$) 
to the in-medium one ($\sim\!1$). But the actual quenching in nuclear structure 
calculations can depend as on the detailed nuclear environment as on the truncations inherent 
to the model such as, for example, the basis size. Therefore, it appears useful to 
revisit the theoretical explanations of quenching, in order to check if and how they can 
cover cases with $g_A<1$, as those emerging from our phenomenological analysis.

From the experimental viewpoint, it has been already mentioned that future EC data 
\cite{TiEC,Frek} will be especially relevant in improving the $(g_{pp},g_A)$ parameter 
constraints.  Moreover, the strong quenching of the axial vector coupling constant $g_A$
should be observed not only in single and double beta decays, but also in M1 transitions. Therefore,
the study of charge-exchange reactions as $(p,\,n)$, $(n,\,p)$, ($^3$He,$\,t$) and ($d,\,^2$He) 
\cite{Ejir,Rake,Am08} can shed new light on this issue. It is imperative, however, that
the data are analyzed with no prior or hidden hypotheses about the GT coupling $g_A$.

In conclusion, we think that 
the results of this work offer a novel possibility to reconcile QRPA results
with experimental data, which deserves further discussions and tests, and warrants 
a revisitation of the quenching problem from a new perspective. 
By the present analysis, we are able to assign in a controlled manner theoretical uncertainties
to the calculated matrix elements for the $0\nu2\beta$ decay. Remarkably, our present results for $M'^{0\nu}$ 
agree within the error bars with those obtained in~\cite{Ro07} for $g_A=1.0$. 
\medskip
\medskip

\ack

This work is supported in part by the EU ILIAS project under the contract RII3-CT-2004-506222. 
The work of G.L.F, E.L., and A.M.R.\ is also supported by the Italian Istituto Nazionale di Fisica 
Nucleare (INFN) and Ministero dell'Universit\`a e Ricerca (MiUR) through the ``Astroparticle Physics'' 
research project. A.F., V.R., and F.\v{S}.\ acknowledge support of the Deutsche Forschungsgemeinschaft 
within the Transregio SFB Project TR27 ``Neutrinos and Beyond'' and by the grant FA67/28-2; in addition, 
F.\v{S}.\ was supported by the VEGA Grant agency of the Slovak Republic (contract No.~1/0249/03)
and by the DFG (436 SLK 17/298).

E.L.\ thanks the Institute of Theoretical Physics (Tuebingen, Germany), where this work was initiated, 
for kind hospitality. Preliminary results of this work were presented by E.L.\ at the $\nu$MASS workshop 
(Genova, Italy, 2007), and by V.R.\ at the 4th ILIAS-IDEA annual meeting
(Paris, France, 2007). We thank Petr~Vogel for early discussions about global experimental constraints 
on QRPA models.


\newpage
\newpage
\newpage
\newpage
\newpage
\newpage
\newpage

\newpage
\newpage
\end{document}